
%

\def\leaderdot{\leaders\hbox to 1 em {\hss.\hss}\hfill}

\dimen0= \parindent                           
\dimen1= \hsize \advance\dimen1 by -\dimen0   

\dimen2=\baselineskip
\def\skiplines#1 { \dimen3=\dimen2 \multiply\dimen3 by #1 \vskip \dimen3}
\def\fullline{\hbox to \fullhsize}

\def\numpage{\baselineskip=24pt\fullline{\the\footline}}

\def\mathcedilla{\vtop{\hbox{c}{\kern0pt\nointerlineskip}
                 {\hbox{$\mkern-2mu \mathchar"0018\mkern-2mu$}}}}

\mathchardef\gq="0060
\mathchardef\dq="0027
\mathchardef\ssmath="19
\mathchardef\aemath="1A
\mathchardef\oemath="1B
\mathchardef\omath="1C
\mathchardef\AEmath="1D
\mathchardef\OEmath="1E
\mathchardef\Omath="1F
\mathchardef\imath="10 
\mathchardef\fmath="0166
\mathchardef\gmath="0167
\mathchardef\vmath="0176


\def\colleft{\strut\kern.3em}
\def\colright{\kern0pt}

\def\figureh{\hbox to}

\def\m@th{\mathsurround=0pt}
\newif\ifdtpt
\def\displ@y{\openup1\jot\m@th
    \everycr{\noalign{\ifdtpt\dt@pfalse
    \vskip-\lineskiplimit \vskip\normallineskiplimit
    \else \penalty\interdisplaylinepenalty \fi}}}
\def\eqalignl#1{\,\vcenter{\openup1\jot\m@th
                \ialign{\strut$\displaystyle{##}$\hfil&
                              $\displaystyle{{}##}$\hfil&
                              $\displaystyle{{}##}$\hfil&
                              $\displaystyle{{}##}$\hfil&
                              $\displaystyle{{}##}$\hfil\crcr#1\crcr}}\,}
\def\eqalignnol#1{\displ@y\tabskip\centering \halign to \displaywidth{
                  $\displaystyle{##}$\hfil\tabskip=0pt &
                  $\displaystyle{{}##}$\hfil\tabskip=0pt &
                  $\displaystyle{{}##}$\hfil\tabskip=0pt &
                  $\displaystyle{{}##}$\hfil\tabskip=0pt &
                  $\displaystyle{{}##}$\hfil\tabskip\centering &
                  \llap{$##$}\tabskip=0pt \crcr#1\crcr}}
\def\leqalignnol#1{\displ@y\tabskip\centering \halign to \displaywidth{
                   $\displaystyle{##}$\hfil\tabskip=0pt &
                   $\displaystyle{{}##}$\hfil\tabskip=0pt &
                   $\displaystyle{{}##}$\hfil\tabskip=0pt &
                   $\displaystyle{{}##}$\hfil\tabskip=0pt &
                   $\displaystyle{{}##}$\hfil\tabskip\centering &
                   \kern-\displaywidth\rlap{$##$}\tabskip=\displaywidth
                   \crcr#1\crcr}}
\def\eqalignc#1{\,\vcenter{\openup1\jot\m@th
                \ialign{\strut\hfil$\displaystyle{##}$\hfil&
                              \hfil$\displaystyle{{}##}$\hfil&
                              \hfil$\displaystyle{{}##}$\hfil&
                              \hfil$\displaystyle{{}##}$\hfil&
                              \hfil$\displaystyle{{}##}$\hfil\crcr#1\crcr}}\,}
\def\eqalignnoc#1{\displ@y\tabskip\centering \halign to \displaywidth{
                  \hfil$\displaystyle{##}$\hfil\tabskip=0pt &
                  \hfil$\displaystyle{{}##}$\hfil\tabskip=0pt &
                  \hfil$\displaystyle{{}##}$\hfil\tabskip=0pt &
                  \hfil$\displaystyle{{}##}$\hfil\tabskip=0pt &
                  \hfil$\displaystyle{{}##}$\hfil\tabskip\centering &
                  \llap{$##$}\tabskip=0pt \crcr#1\crcr}}
\def\leqalignnoc#1{\displ@y\tabskip\centering \halign to \displaywidth{
                  \hfil$\displaystyle{##}$\hfil\tabskip=0pt &
                  \hfil$\displaystyle{{}##}$\hfil\tabskip=0pt &
                  \hfil$\displaystyle{{}##}$\hfil\tabskip=0pt &
                  \hfil$\displaystyle{{}##}$\hfil\tabskip=0pt &
                  \hfil$\displaystyle{{}##}$\hfil\tabskip\centering &
                  \kern-\displaywidth\rlap{$##$}\tabskip=\displaywidth
                  \crcr#1\crcr}}

\def\doubleup#1{\,\vbox{\ialign{\hfil$##$\hfil\crcr
                  \mathstrut #1 \crcr }}\,}
\def\doublelow#1{\,\vtop{\ialign{\hfil$##$\hfil\crcr
                 \mathstrut #1 \crcr}}\,}
\def\charlvmidlw#1#2{\,\vtop{\ialign{##\crcr
      #1\crcr\noalign{\kern1pt\nointerlineskip}
      $\hfil#2\hfil$\crcr}}\,}
\def\charlvlowlw#1#2{\,\vtop{\ialign{##\crcr
      $\hfil#1\hfil$\crcr\noalign{\kern1pt\nointerlineskip}
      #2\crcr}}\,}
\def\charlvmidup#1#2{\,\vbox{\ialign{##\crcr
      $\hfil#1\hfil$\crcr\noalign{\kern1pt\nointerlineskip}
      #2\crcr}}\,}
\def\charlvupup#1#2{\,\vbox{\ialign{##\crcr
      #1\crcr\noalign{\kern1pt\nointerlineskip}
      $\hfil#2\hfil$\crcr}}\,}

\def\vspce{\kern4pt} \def\hspce{\kern4pt}    

\def\emptybox{\vbox{\kern.7ex\hbox{\kern.5em}\kern.7ex}}
 \font\sevmi  = cmmi7              
    \skewchar\sevmi ='177          
 \font\fivmi  = cmmi5              
    \skewchar\fivmi ='177          
\font\tenmib=cmmib10
\newfam\bfmitfam

\textfont\bfmitfam=\tenmib
\scriptfont\bfmitfam=\sevmi
\scriptscriptfont\bfmitfam=\fivmi


\def\twodot{.\kern-0.1em.}

\def\paral{\mathrel{/\kern-.25em/}}
\def\grlo{\mathrel{\hbox{\lower.2ex\hbox{\rlap{$>$}\raise1ex\hbox{$<$}}}}}
\def\logr{\mathrel{\hbox{\lower.2ex\hbox{\rlap{$<$}\raise1ex\hbox{$>$}}}}}
\def\greq{\mathrel{\hbox{\lower1ex\hbox{\rlap{$=$}\raise1.2ex\hbox{$>$}}}}}
\def\loeq{\mathrel{\hbox{\lower1ex\hbox{\rlap{$=$}\raise1.2ex\hbox{$<$}}}}}
\def\grsim{\mathrel{\hbox{\lower1ex\hbox{\rlap{$\sim$}\raise1ex\hbox{$>$}}}}}
\def\losim{\mathrel{\hbox{\lower1ex\hbox{\rlap{$\sim$}\raise1ex\hbox{$<$}}}}}
\font\ninerm=cmr9
\def\uniset{\rlap{\ninerm 1}\kern.15em 1}

\def\emptysq{\mathbin{\vbox{\hrule\hbox{\vrule height1ex \kern.5em 
                            \vrule height1ex}\hrule}}}
\def\emptyrect{\mathbin{\vbox{\hrule\hbox{\vrule height1ex \kern1em 
                              \vrule height1ex}\hrule}}}
\def\rightonleftarrow{\mathrel{\hbox{\raise.5ex\hbox{$\rightarrow$}\ignorespaces
                                   \lower.5ex\hbox{\llap{$\leftarrow$}}}}}
\def\leftonrightarrow{\mathrel{\hbox{\raise.5ex\hbox{$\leftarrow$}\ignorespaces
                                   \lower.5ex\hbox{\llap{$\rightarrow$}}}}}

\def\bkB{{\rm I\kern-.17em B}}
\def\bkC{{\rm \kern.24em
            \vrule width.05em height1.4ex depth-.05ex
            \kern-.26em C}}
\def\bkD{{\rm I\kern-.17em D}}
\def\bkE{{\rm I\kern-.17em E}}
\def\bkF{{\rm I\kern-.17em F}}
\def\bkG{{\rm \kern.24em
            \vrule width.05em height1.4ex depth-.05ex
            \kern-.26em G}}
\def\bkH{{\rm I\kern-.22em H}}
\def\bkI{{\rm I\kern-.22em I}}
\def\bkJ{{\rm \kern.19em
            \vrule width.02em height1.5ex depth0ex
            \kern-.20em J}}
\def\bkK{{\rm I\kern-.22em K}}
\def\bkL{{\rm I\kern-.17em L}}
\def\bkM{{\rm I\kern-.22em M}}
\def\bkN{{\rm I\kern-.20em N}}
\def\bkO{{\rm \kern.24em
            \vrule width.05em height1.4ex depth-.05ex
            \kern-.26em O}}
\def\bkP{{\rm I\kern-.17em P}}
\def\bkQ{{\rm \kern.24em
            \vrule width.05em height1.4ex depth-.05ex
            \kern-.26em Q}}
\def\bkR{{\rm I\kern-.17em R}}
\def\bkT{{\rm \kern.24em
            \vrule width.02em height1.5ex depth 0ex
            \kern-.27em T}}
\def\bkU{{\rm \kern.30em
            \vrule width.02em height1.47ex depth-.05ex
            \kern-.32em U}}
\def\bkZ{{\rm Z\kern-.32em Z}}



\magnification=1200
\centerline{MIXED ELLIPTIC AND HYPERBOLIC SYSTEMS FOR THE EINSTEIN EQUATIONS}
\smallskip
\centerline{Yvonne Choquet-Bruhat and James W. York}
\medskip
\noindent\underbar{INTRODUCTION.} \par
\noindent We examine the Cauchy problem$ ^5 $ for General Relativity as the
time history 
of the two fundamental forms of the geometry of a space like hypersurface, 
its metric $ \bar  g $ and its extrinsic curvature K. By using a 3+1
decomposition of 
the Ricci tensor, we split$^{1,2}$ the Einstein equations 
into ``constraints", 
equations containing only $ \bar  g $, K and their space derivatives, and
evolution 
equations giving the time derivatives of $ \bar  g $ and K in terms of space 
derivatives of these quantities and also of the lapse and shift, i.e., the 
choice of the time lines. The constraints can be posed and solved as an 
elliptic system by known methods. However the equations of evolution in 
time for $ \bar  g $ and K are not, despite their form$ ^3 $ as canonical
hamiltonian 
equations, a mathematically well posed system; and they do not manifest 
directly the 
propagation of physical gravitational effects along the light cone. These 
equations, of course, contain gauge effects and cannot, therefore, yield 
directly a physical wave equation, that is, a hyperbolic system with 
suitable characteristics. (For a review of the relevant geometry, see, for 
example$ ^9 $.) \par
\noindent In this paper we give two different methods for obtaining a
nonlinear wave 
equation for the evolution of the extrinsic curvature K. The evolution of $
\bar  g $ 
is just its dragging by K along the axis orthogonal to the time slices. \par
\noindent The first method corresponds to a choice of time slicing
 by fixation of the mean
extrinsic 
curvature of the space slices. It leads to a mixed elliptic and hyperbolic 
system for which we prove local in time, global in space, existence theorems 
in the cases of compact or asymptotically euclidean space slices. \par
\noindent The second method, which relies on a harmonic time slicing
condition, 
leads to equations of motion equivalent to a first order symmetric 
hyperbolic system with only physically appropiate characteristics. We 
construct this system explicitly. Among the propagated quantities is, in 
effect, the Riemann curvature. The space coordinates and the shift are 
arbitrary; and, in this sense, the system is gauge invariant. \par
\noindent Our gauge invariant non linear hyperbolic system, being
exact and always 
on the physical light cone, is well suited for use in a number of problems 
which now confront gravity theorists. These include large scale computations 
of astrophysically significant processes (such as black hole collisions) 
that require efficient stable numerical integration, extraction of 
gravitational radiation with arbitrarily high accuracy from Cauchy data, 
gauge invariant perturbation and approximation methods, and posing boundary 
conditions compatibly with the causal structure of space time. \par
\smallskip
\noindent n+1 DECOMPOSITION, ARBITRARY SHIFT. \par
\smallskip
\noindent The n+1 decomposition of the curvature of a lorentzian manifold and
the 
resulting decomposition of Einstein equations into constraints and 
evolution equations have been given in the case of zero shift by 
Lichnerowicz (1939) and in the case of an arbitrary shift by Choquet 
(Foures)- Bruhat (1956). Arnowitt, Deser and Misner (1962) have obtained 
this decomposition through a hamiltonian formalism and have deduced from 
it interesting physical consequences, for example, the definition of the 
gravitational mass. \par
\noindent We recall this straightforward decomposition, using a coframe with
time 
axis orthogonal to the space slices and introducing the notation of spatial 
Lie derivative like, for example, Fischer and Marsden$^4$. We consider on a 
manifold V 
= M$\times \bkR$ a pseudo riemannian metric of lorentzian signature which
reads \par
\smallskip
\centerline{$ ds^2 \equiv  g_{\alpha \beta} \theta^ \alpha \theta^ \beta 
\equiv  -N^2(\theta^ 0)^2 + g_{ij}\theta^ i\theta^ j $}
\smallskip
\noindent with (t $\in$ $\bkR$, $ x^i $, i = 1,..,n are local coordinates on
M) \par
\smallskip
\centerline{$ \theta^ 0 \equiv  dt\ \ \ \ \ \ \ \ ,\ \ \ \ \ \ \ \ \theta^ i
\equiv  dx^i + \beta^ idt $}
\smallskip
\noindent With this choice of coframe, the dual frame has a time axis
orthogonal to 
the space slices $ M_t $ $ \equiv  M\times \lbrace t\rbrace   $ while the
space axis are tangent to them. The 
Pfaff derivatives $ \partial_ \alpha $ with respect to $\theta$ $ ^\alpha $
are \par
\smallskip
\centerline{$ \partial_ 0 \equiv  {\partial \over \partial t} - \beta^
i\partial_ i\ \ \ \ \ \ ,\ \ \ \ \ \ \partial_ i \equiv  {\partial \over
\partial x^i} $}
\smallskip
\noindent The structure constants of our coframe are such that \par
\smallskip
\centerline{$ d\theta^ \alpha  = - {1 \over 2} c^\alpha_{ \beta \gamma} 
\theta^ \beta\wedge \theta^ \gamma $}
\smallskip
\noindent Hence \par
\smallskip
\centerline{$ c^0_{\alpha \beta}  = 0\ \ \ ,\ \ \ c^i_{jk} = 0\ \ \ ,\ \ \
c^i_{j0} = - c^i_{0j} = - \partial_ j\beta^ i $}
\smallskip
\noindent We denote by  $ \bar  \Gamma\allowbreak^ i_{jk} $ the Christoffel
symbols of the (time dependent) space 
metric $ \bar  g $ $\equiv$ g$ _{ij}dx^idx^j $, and by 
$ \omega^ \lambda_{ \alpha \beta} $ 
the coefficients of the riemannian connection 
of g in the coframe $ \theta^ \alpha $. We have \par
\smallskip
\centerline{$ \omega^ i_{jk} \ =\ \bar  \Gamma^ i_{jk} $}
\smallskip
\centerline{$ \omega^ 0_{00} = N^{-1}\partial_ 0N\ \ \ ,\ \ \ \omega^ 0_{0i} =
\omega^ 0_{i0} = N^{-1}\partial_ iN $}
\smallskip
\centerline{$ \omega^ 0_{ij} = {1 \over 2} N^{-2}(\partial_ 0g_{ij} - g_{h(i}
\partial_{ j)}\beta^ h) \equiv  {1 \over 2} N^{-2}({\partial \over \partial t}
g_{ij} - {\cal L}_\beta g_{ij} $)}
\smallskip
\noindent where we define the symmetrization by $ f_{(ij) } $$\equiv$ $ f_{ij}
+ f_{ji} $ (no factor of 1/2). \par
\noindent We define for any t dependent space tensor T another such tensor $
\hat  \partial_ 0T $ of the same 
type by setting \par
\smallskip
\centerline{$ \hat  \partial\allowbreak_ 0  \equiv  {\partial \over
\partial}_ t- {\cal L}_\beta $}
\smallskip
\noindent where $ {\cal L}_\beta $ is the Lie derivative on $ M_t $ with
respect to $\beta$. \par
\noindent The extrinsic curvature of the space slices is given by \par
\smallskip
\centerline{$ K_{ij} \equiv  - N \omega^ 0_{ij} \equiv   $- $ {1 \over 2}
N^{-1} \hat  \partial_ 0g_{ij} $}
\smallskip
\noindent and we find \par
\smallskip
\centerline{$ \omega^ k_{0i} = -N K^k_i\ \ \ ,\ \ \ \omega^ k_{i0} = - N K^k_i
+ \partial_ i\beta^ k $}
\smallskip
\noindent\underbar{Curvature tensor} \par
\noindent The components of the curvature tensor of g in the frame $ \theta^
\alpha $ are found to 
be, using the general formulas given in Choquet-Bruhat and DeWitt-Morette 
I,  p. 306, (with an opposite sign) \par
\smallskip
\centerline{$ R^{\ j}_{i\ kl} = \bar  R^{\ j}_{i\ kl} - K^j_k K_{li} + K^j_l
K_{ki} $}
\smallskip
\centerline{$ R^{\ 0}_{i\ kl} =  N^{-1}(\bar  \nabla_ kK_{li} - \bar  \nabla_
lK_{ki} $)}
\smallskip
\centerline{$ R^{\ 0}_{i\ j0} = - ( N^{-1} \hat  \partial_ 0K_{ij} +
K_{im}K^m_j + N^{-1} \bar  \nabla_ j\partial_ iN $ )}
\smallskip
\noindent\underbar{Ricci tensor.} \par
\noindent From these formulas result the following ones for the Ricci
curvature, 
where we have denoted by H the mean extrinsic curvature of the space 
slices, i.e., we have set
\par
\centerline{H $\equiv$ $ K^h_h $,}
\smallskip
\centerline{$ R_{ij} \equiv  - N ^{-1} \hat  \partial_ 0K_{ij} + H K_{ij} -
2K_{im}K^m_j - N^{-1} \bar  \nabla_ j\partial_ iN + \bar  R_{ij}  $}
\smallskip
\centerline{$ R^{ }_{0j} \equiv  N^{ }(- \bar  \nabla_ hK^h_j + \partial_ jH)
$}
\smallskip
\centerline{$ R^{ }_{00} \equiv  N (\bar  \nabla^ i\partial_ iN - N
K_{ij}K^{ij} + \doubleup{ \cr \partial \cr}_ 0H $)}
\medskip
\noindent SECOND ORDER EQUATION FOR K. \par
\smallskip
\noindent In the formula above R$ _{ij} $ is, like the right hand side giving
its 
decomposition, a t dependent space tensor, the projection on space of the Ricci 
tensor of the space time metric. We compute its $ \hat  \partial_ 0 $
derivative. First we 
compute $ \hat  \partial_ 0\bar  R_{ij}  $. The infinitesimal variation of the
Ricci curvature 
corresponding to an infinitesimal $\delta \bar  g $ variation of the metric
is \par
\smallskip
\centerline{$ \delta\bar  R_{ij} =  {1 \over 2} \lbrace\bar  \nabla^ h\bar 
\nabla_{ (i}\delta g_{j)h}  -  \bar  \nabla_ h $ $ \bar  \nabla^ h \delta
g_{ij} - \bar  \nabla_ j\partial_ i(g^{hk}\delta g_{hk}) $$\rbrace$ }
\smallskip
\noindent This expression applies to ($\partial$/$\partial$t)$ \bar
 R _{ ij} $ with $\delta g_{ij} = (\partial /\partial t)g_{ij} $ and
to ${\cal L} _\beta \bar R _{ ij} $ 
with $\delta g_{ij} = {\cal L}_\beta g_{ij} $. Therefore, using the
relation between $ \hat  \partial_ 0g_{ij} $ and $K_{ij} $, we 
obtain \par
\smallskip
\centerline{$ \hat  \partial_ 0\bar  R_{ij} \equiv  \allowbreak - \bar 
\nabla^ h\bar  \nabla_{ (i}(NK_{j)h} $) + $ \bar  \nabla_ h\bar  \nabla^
h(NK_{ij}) + \bar  \nabla_ j\partial_ i(NH) $}
\medskip
$ \equiv  \allowbreak - \bar  \nabla_{ (i}(N\bar  \nabla^ h(K_{j)h}) -  \bar 
\nabla_{(i}(K_{j)}^h\partial _h N - 2 N \bar  R\allowbreak^ h_{\ ijm}K^m_h -
N\bar  R_{m(i}K^m_{j)} $ \par
\smallskip
+ $ \bar  \nabla_ h\bar  \nabla^ h(NK_{ij}) + \bar  \nabla_ j\partial_ i(NH)
$. \par
\smallskip
\noindent We now use the expressions for $ R^{ }_{0i} $ and $ R_{ij} $ 
to obtain the identity \par
\smallskip
\noindent$ \Omega_{ ij} \equiv $ $ \hat  \partial_ 0R_{ij}  -  \bar
 \nabla _{ (i}R^{ }_{j)0} \equiv   $ \par
\smallskip
\noindent$  -  \hat  \partial_ 0(N ^{-1} \hat  \partial_ 0K_{ij} ) +  \hat 
\partial_ 0(H K_{ij} - 2K_{im}K^m_j) - \hat  \partial_ 0(N^{-1} \bar
 \nabla _ j\partial_ iN) - N \bar \nabla i\partial_
jH $ \par
\smallskip
\centerline{$- \bar \nabla _{ (i}(K_{j)h}\allowbreak
\partial^ hN ) - 2N \bar R ^ h_{\ ijm}K^m_h - N \bar R _{ m(i}K^m_{j)} $ 
+ $ \bar \nabla _ h
\bar \nabla ^ h(NK_{ij}) + H \bar \nabla _
j\partial_ iN $}
\medskip
\noindent This identity shows that for a solution of the Einstein equations
\smallskip
$$ R_{\alpha \beta}  = \rho_{ \alpha \beta} \leqno (E) $$
\par
\noindent the extrinsic curvature $ K $ satisfies a second order differential
system 
which is quasidiagonal with principal part the wave operator, except for 
the terms $ \bar  \nabla_ i\partial_ jH$.   The other unknowns $ \bar  g $ and
N appear at second order except 
for the term $ \hat  \partial_ 0\bar  \nabla_ j\partial_ iN $ . We will use
these facts in the following paragraphs 
to obtain geometrical well posed systems for the Einstein equations.

\bigskip
\noindent I. MIXED HYPERBOLIC AND ELLIPTIC SYSTEM FOR $ \bar  g $, K, N WHEN H
IS GIVEN. \par
\smallskip
\noindent A procedure to reduce the second order equation for K obtained above
to a 
quasi diagonal system with principal part the wave operator is to replace 
in the term $ \bar  \nabla_ j\partial_ iH$  the mean curvature H by an priori
given function h: this 
is a gauge condition on the space slices. It was used by Christodoulou and 
Klainerman$^7$, with h = 0, in the asymptotically euclidean case. With this 
replacement the equations $\Omega_{ij} = \Theta_{ ij} $ become, when N is
known, a quasi 
diagonal second order system for K with principal part the wave operator, 
namely: \par
\smallskip
\centerline{$ \emptysq K_{ij} = P_{ij} + \Theta_{ ij} $}
\smallskip
\noindent with \par
\centerline{$  \emptysq K_{ij}\equiv   -  \hat  \partial_ 0 (N^{-1} \hat 
\partial_ 0K_{ij}) + \bar  \nabla^ h\bar  \nabla_ h(NK_{ij}) $}
\noindent where $ P_{ij}  $ depends only on $ K $ and its first derivatives,
on $ \bar  g $, N and $\partial_0 $N 
together with their space derivatives of order $\leq$ 2, and is given by: \par
\medskip
\centerline{$ \eqalignc{ P_{ij}\equiv   \hat  \partial_ 0(- H K_{ij} + 2
g^{hm}K_{ih} K_{jm)}) +  \hat  \partial_ 0(N^{-1}\bar  \nabla_ j\partial_
iN) \cr} $}
\medskip
+ $ \bar  \nabla_{(i} (K_{j)h} \partial ^hN) + 2N \bar  R^h_{\ ijm}K^m_h +
N\bar  R_{m(i}K^m_{j)}  -   H \bar  \nabla_ j\partial_ iN  +  N \bar 
\nabla_ j\partial_ ih $ \par
\medskip
\noindent while $ \Theta_{ ij} $, zero in vacuum, is: \par
\smallskip
\centerline{$ \Theta_{ ij} \equiv\allowbreak   \hat  \partial_ 0\rho_{ ij} - 
 \bar  \nabla_{ (i}(\rho _{j)0})   $}
\smallskip
\noindent When $\beta$, N and the sources $\rho$ are known, the above equation
together with \par
\smallskip
$$  \hat \partial_ 0g_{ij} = -2N K_{ij} \leqno(g') $$ \par
\smallskip
\noindent constitute a third order quasi diagonal system for $ \bar  g $,
hyperbolic if $N^2 >$ 0 
and $ \bar  g $ is properly riemannian. \par
\noindent On the other hand the equation $ R^0_0 = \rho^ 0_0 $ together with H
= h imply the 
equation \par
\smallskip
\centerline{$  \bar  \nabla^ i\partial_ iN - (K_{ij}K^{ij} - \rho^ 0_0) N = -
 \partial _ 0h $}
\smallskip
\noindent This equation is an elliptic equation for N when $ \bar  g $ , K and
$\rho$ are known. \par
\noindent Note that for energy sources satisfying the energy condition we have
$  - \rho^ 0_0 \geq  0$  as well as $K.K  \equiv  K_{ij}K^{ij} $
$\geq$ 0, an important property for the 
solution of the elliptic equation. The mixed hyperbolic elliptic system 
that we have constructed will determine the unknowns N and $ \bar  g $ in a 
neighbourhood of M in M$\times \bkR$ when the shift $\beta$ is chosen. \par
\medskip
\noindent LOCAL EXISTENCE THEOREM(case of given H). \par
\smallskip
\noindent We consider the \underbar{vacuum} case. The system for $ \bar  g, K
$ \par
\smallskip
\noindent$ (1)\ \ \ \ \ \ \ \ \ \ \ \ \ \ \ \ \ \ \ \ \hat  \partial_ 0g_{ij}
= - 2N K_{ij}  $ \par
\smallskip
\noindent (2)\nobreak\ \nobreak\ \nobreak\ \nobreak\ \nobreak\ \nobreak\
\nobreak\ \nobreak\ \nobreak\ \nobreak\ \nobreak\ \nobreak\ \nobreak\
\nobreak\ \nobreak\ \nobreak\ \nobreak\ \nobreak\ \nobreak\ \nobreak\
\nobreak\ \nobreak\ $ \emptysq K\allowbreak_{ ij} = P_{ij} $ \par
\smallskip
\noindent is equivalent to a third order hyperbolic system. It requires for
its 
solution only local Sobolev spaces (cf. Leray$ ^8$), but the lapse N is now 
determined by an elliptic equation on each $ M_t $ which must be solved 
globally: \par
\smallskip
\noindent$ (3)\ \ \ \ \ \ \ \ \ \ \ \ \ \ \ \bar  \nabla  ^i\partial_ iN - K.K
N = - \partial_ 0h $ \par
\smallskip
\noindent We will consider the two cases of a compact and of an asymptotically
euclidean M. In both cases we will use an iteration scheme solving 
alternatively the elliptic equation and a linear hyperbolic system deduced 
in an obvious manner from (1), (2). \par
\noindent We specify the results in the physical case of \underbar{space
dimensions n = 3}.  
There are problems in extending them to other dimensions in the asymptotically 
euclidean case. \par
\noindent{\bf Case of a compact M.} \par
\smallskip
\noindent In the case of a compact M a good choice in order to have a positive
lapse 
is to take $ \partial_ 0h = f > 0. $ Standard elliptic theory gives the
following lemma, 
where Sobolev spaces $ H_s $ are relative to some smooth fixed metric e on M,
I 
is an interval of $\bkR$ and the shift $\beta$ is taken smooth on M$\times$I.
\par
\smallskip
\noindent\underbar{Lemma 1.} a) If $ \bar  g(t) \in  H_3, K(t)
\in  H_2 $, with K.K $\not\equiv$ 0, and f $\in$ H$ _2  $ the lapse 
equation has a unique solution N(t) $\in$ $ H_4 $. By the maximum principle
this 
solution is such that N $>$ 0 on M if f $\geq$ 0 on M and f $\not\equiv$ 0.
\par
\noindent b) If moreover $ \bar  g \in  \doublelow{ \cap \cr 1\leq k\leq 3
\cr}  C^{3-k}(I,H_k) $, while f, $ K \in   \doublelow{ \cap \cr 0\leq k\leq 2
\cr} C^{2-k}(I,H_k)  $ \par
\noindent then \par
\centerline{N $\in$ $ \doublelow{ \cap \cr 0\leq k\leq 2 \cr} $ $
C^{2-k}(I,H_{2+k}). $}
\medskip
\noindent\underbar{Remark.} We have $ K.K \geq {1\over 3} H^ 2 $ hence
$ K.K$ $ \not\equiv $ 0 if $ H $ $\not\equiv$ 0; we will prove later 
that the equality $ H = h $ holds for a solution of the coupled hyperbolic and
elliptic system if it holds initially, but we cannot use this property in 
the iteration scheme. The condition $ K.K$ $ \not\equiv $ 0  must be deduced from such
an 
hypothesis on the initial data and the continuity properties possessed by 
solutions of the hyperbolic system. \par
\medskip
\noindent We now consider the linear system for $ \bar  g $(n+1), $ K $(n+1)
\par
\smallskip
\noindent$ (1_n)\ \ \ \ \ \ \ \ \ \ \ \ \ \ \ \ \ \ \ \ \hat  \partial_
0g_{ij}(n+1) = - 2N(n) K_{ij}(n)  $,\nobreak\ \nobreak\ \nobreak\ \par
\smallskip
\noindent (2$ _n $)\nobreak\ \nobreak\ \nobreak\ \nobreak\ \nobreak\
\nobreak\ \nobreak\ \nobreak\ \nobreak\ \nobreak\ \nobreak\ \nobreak\
\nobreak\ \nobreak\ \nobreak\ \nobreak\ \nobreak\ \nobreak\ \nobreak\
\nobreak\ \nobreak\ \nobreak\ $ \emptysq\allowbreak_ n $K$ _{ij}(n+1) = P_{ij}
$(n) \par
\smallskip
\noindent obtained by replacing in $\emptysq$ the metric $ \bar  g $ by a
metric $ \bar  g $(n) and in $ P_{ij} $ also $ K $ 
by $ K $(n) and N by N(n). \par
\smallskip
\noindent\underbar{Lemma 2.} \underbar{Hypothesis.}  Given: \par
\noindent 1)a)\nobreak\ \nobreak\  N(n) $\geq$ N$ _0 $ $>$ 0 , N(n) $\in$ $ 
\doublelow{ \cap \cr 0\leq k\leq 2 \cr} C^{2-k}(I,H_{2+k}) $, \par
\noindent b)\nobreak\ \nobreak\ h $\in$ $ \doublelow{ \cap \cr 2\leq k\leq 3
\cr} $ $ C^{3-k}(I,H_k), I \equiv  \lbrack 0,T\rbrack $ , \par
\noindent c)\nobreak\ \nobreak\ $ \bar  g(n) \in   \doublelow{ \cap \cr 1\leq
k\leq 3 \cr} C^{3-k}(I,H_k) \ \ {\rm and} \ \ K(n) \in   \doublelow{ \cap \cr
0\leq k\leq 2 \cr} C^{2-k}(I,H_k) $. \par
\noindent 2) Cauchy data such that \par
\smallskip
\centerline{$ g_{ij}(n+1)(0,.) = \gamma_{ ij} \in  H_3,\ \ \ K_{ij}(n+1)(0,.)
= k_{ij} \in  H_2 $,}
\smallskip
\noindent where the metric $\gamma$ is uniformly equivalent to the given metric e
and  $ k.k $ $\not\equiv$ 0 on 
M. \par
\noindent The data $ \partial_ 0K_{ij}(n+1)(0,.) = \dot  k_{ij} $, determined
by the equation $ R_{ij}(0,.) = 0 $, 
belong to $ H_1 $. \par
\smallskip
\noindent\underbar{Conclusion.} The Cauchy problem for the system (1$ _n $),
(2$ _n $) has a unique 
solution \par
\smallskip
\centerline{$ \bar  g(n+1) \in   \doublelow{ \cap \cr 1\leq k\leq 3 \cr}
C^{3-k}(I,H_k),\ \ \  K(n+1) \in   \doublelow{ \cap \cr 0\leq k\leq 2 \cr}
C^{2-k}(I,H_k) $.}
\medskip
\noindent Moreover there exists an interval $ I_1= \lbrack 0,T_1\rbrack $, 0
$<$ T$ _1 $ $\leq$ T, and a number $\epsilon >$ 0  
depending only on the norms of the given quantities in their respective 
spaces such that $ \bar  g $(n+1) is uniformly equivalent to $\gamma$ on
M$\times$ $ I_1 $ and 
\par
\centerline{$ \parallel K(n+1).K(n+1)\parallel_{ C^0(M\times I_1)} $ $\geq$
$\epsilon$ .}
\medskip
\noindent\underbar{Proof.} The system is equivalent to a third order
hyperbolic system, and the 
results can be proved by standard methods. \par
\smallskip
\noindent\underbar{Remark}. When $ \bar  g $(n), $ K $(n), N(n) satisfy the
indicated hypothesis the equation 
(2$ _n$)  is a wave equation for K(n+1) whose coefficients satisfy the
required 
hypothesis for the existence of a solution in $ \doublelow{ \cap \cr 0\leq
k\leq 2 \cr} $C$ ^{2-k}(I,H_k) $. The 
integration of equation (1$ _n $) determines then $ \bar  g $(n+1), but the
general theory 
of first order equations is not sufficient to assert its required 
regularity. One must use the fact that $ \bar  g(n+1) $ satisfies a third
order 
hyperbolic equation to prove that result. \par
\noindent We remark also that the regularity hypothesis made on h is stronger
that 
the regularity we prove on $ K $. \par
\smallskip
\noindent{\bf Theorem.} Let (M,e) be a smooth compact riemannian manifold. Let
be given on 
M$\times$I, with I $\equiv$ $\lbrack$0,T$\rbrack$, a pure space smooth
vector field $\beta$ and a function h such that \par
\smallskip
\centerline{h $\in$ $ \doublelow{ \cap \cr 2\leq k\leq 3 \cr} $ $
C^{3-k}(I,H_k) $ ,\nobreak\ \nobreak\ \nobreak\ $ \partial_ 0h $ $\geq$ 0, $
\partial_ 0h $ $\not\equiv$ 0 .}
\smallskip
\noindent There exists an interval J $\equiv$ $\lbrack$0,$\ell \rbrack$,
$\ell$ $\leq$ T such that the system (1), (2), (3) 
has one and only one solution on M$\times$J \par
\smallskip
\noindent$ \bar  g \in   \doublelow{ \cap \cr 1\leq k\leq 3 \cr}
C^{3-k}(J,H_k), K \in   \doublelow{ \cap \cr 1\leq k\leq 2 \cr} C^{2-k}(J,H_k)
$, N $\in$ $ \doublelow{ \cap \cr 0\leq k\leq 2 \cr} $C$ ^{2-k} $(J,H$ _{2+k}
$). \par
\smallskip
\noindent with N $>$ 0 and $ \bar  g $ uniformly equivalent to e, taking the
initial data \par
\smallskip
\centerline{$ g_{ij}(0,.) = \gamma_{ ij} \in  H_3,\ \ \ K_{ij}(0,.) = k_{ij}
\in  H_2 $}
\smallskip
\noindent if $\gamma$ is a properly riemannian metric uniformly equivalent to
e and k.k $\not\equiv$ 0. \par
\smallskip
\noindent\underbar{Proof}. One determines N(0) on M$\times$I by the equation
on each M$ _t $ \par
\smallskip
\centerline{$\bigtriangleup$ $ _\gamma N(0) - k.k N(0) = -\partial_ 0h(0) $}
\smallskip
\noindent and then proceeds to the iterations, with N(n) determined from $
\bar  g $(n), $ K $(n). \par
\noindent One uses the elliptic estimates for N(n) and the energy estimates of
first 
and second order for $ K $(n) and $ \bar  g(n) $ to prove the boundedness of
the iterates 
in\nobreak\ the indicated norms, as well as to show that there exists $ I_1
=\lbrack 0,\ell_ 1\rbrack $ $\subset$ I such 
that, on M$\times$I$ _1 $, $ \bar  g(n) $ is uniformly equivalent to e and
$\parallel$ $ K $(n).$ K $(n)$\parallel$ $ _{C^0 }\geq  \epsilon  > 0 $. \par
\noindent The elliptic estimates and the first order energy estimates applied
to the 
difference of two successive iterates show that there exists J $\equiv$
$\lbrack$0,$\ell \rbrack$ 
such that the sequence converges strongly but in a weaker norm 
\par
\noindent$  (C^0(J,H_2)\cap C^1(J,H_1))\times  (C^0(J,H_1)\cap C^1(J,H_0)
$)$\times$ (C$ ^0(J,H_3)\cap C^1(J,H_2) $). \par
\smallskip
\noindent Since the sequence is bounded in $ L^{\infty} (J,H_3)\times
L^{\infty} (J,H_2)\times L^{\infty} (J,H_4)$, the dual of a 
Banach space, there exists a subsequence which converges weakly in that 
space, and the limit found from the strong convergence belongs therefore to 
that space. An analogous argument applied to the time derivatives shows the 
existence of a subsequence whose first $\lbrack$respectively second$\rbrack$
time 
derivatives converge weakly in L$ ^{\infty} (J\times H_2)\times L^{\infty}
(J\times H_1) $ $\lbrack$respectively in L$ ^{\infty} (J,H_1) $ 
$\times$ $ L^{\infty} (J\times H_0) $$\rbrack$, and the limit found belongs
therefore to these spaces. It can be 
proved using properties of strong and weak convergence that the limit 
satisfies the non linear system (1), (2), (3). It can also be proved that 
this solution is in fact in the space given in the theorem (cf. a proof for 
second order systems by Hughes, Kato and Marsden$^{10}$ and for
symmetric 
hyperbolic first order systems by Majda$^{11}$). \par
\medskip
\noindent{\bf Case of (M,e) euclidean at infinity.} \par
\smallskip
\noindent We now consider the case of a manifold M which is the union of a
compact 
set and a finite number of disjoint sets (``ends") diffeomorphic to the 
exterior of a ball in $\bkR^3 $; M is endowed with a smooth properly
riemannian 
metric e which reduces on each end to the euclidean metric. \par
\noindent We will look for an asymptotically minkowskian solution of (1), (2),
(3) 
on M$\times$I, i.e. a solution belonging to some weighted Sobolev space $
H_{s,\delta} $ on 
each space slice M$ _t $. We recall that $ H_{s,\delta} $ is the completion of
C$ ^{\infty}_ 0 $ in the 
norm \par
\smallskip
\centerline{$\parallel$f$\parallel$ $ _{H_{s,\delta}} $ $\equiv$ $\lbrace$ $
\int^{ }_ M \sum^{ }_{  0\leq k\leq s}\sigma^{ 2k+2\delta} \mid D^kf\mid^
2\mu (e)\rbrace^{ 1/2} $}
\smallskip
\noindent where $\sigma$ $ ^2 $ $\equiv$ 1+d$ ^2 $, d the distance in the
metric e to some fixed point in M. \par
\noindent The theory of elliptic equations on an asymptotically euclidean
manifold in 
$ H_{s,\delta}   ${\bf \nobreak\ }spaces (Choquet-Bruhat and Christodolou$^{12}$)
gives the following lemma. \par
\smallskip
\noindent\underbar{Lemma 1.} a) If $ \bar  g(t) - e \in  H_{3,-1},
K(t) \in  H_{2,0} $ and f $\equiv$ $ \partial_ 0h $ $\in$ H$ _{2,1}  ${\bf
\nobreak\ }the lapse 
equation has a unique solution with 1 - N(t) $\in$ $ H_{4,-1} $. By the
maximum 
principle this solution is such that N $>$ 0 on M if $ \partial_ 0h $ $\geq$ 0
on M. \par
\noindent b) If moreover: \par
\smallskip
\noindent$ \bar  g - e \in  \doublelow{ \cap \cr 1\leq k\leq 3 \cr} 
C^{3-k}(I,H_{k,-1}) $,\nobreak\ \nobreak\ \nobreak\ \nobreak\  $  K \in  
\doublelow{ \cap \cr 0\leq k\leq 2 \cr} C^{2-k}(I,H_{k,0}) $, \par
\smallskip
\noindent f $ \in  \doublelow{ \cap \cr 0\leq k\leq 2 \cr}\allowbreak
C^{2-k}(I,H_{k,1} $) \par
\smallskip
\noindent then \par
\smallskip
\noindent 1 - N $\in$ $ \doublelow{ \cap \cr 0\leq k\leq 2 \cr} $ $
C^{2-k}(I,H_{2+k,-1}). $ \par
\smallskip
\noindent\underbar{Proof.} a) The multiplication properties of weighted
Sobolev spaces show that 
with the hypothesis made $ (K.K $)(t) $\in$ $ H_{2,1} $. Since - $ {3 \over 2}
$ $<$ -1 $<$ - $ {1 \over 2} $ the equation \par
\smallskip
\noindent$ \ \ \ \ \ \ \ \ \ \ \ \ \ \ \ \bar  \nabla  ^i\partial_ iu - K.K u
= - \partial_ 0h $ + $ K.K $ ,\nobreak\ \nobreak\ \nobreak\ \nobreak\
\nobreak\ \nobreak\ N = 1 + u \par
\smallskip
\noindent has one and only one solution u $\in$ H$ _{4,-1} $; one deduces the
property N $>$ 0 
from the maximum priciple and the fact that N tends to 1 at infinity . \par
\noindent b) The first time derivative of u can be proven to be in $
H_{3,\delta} $, but again 
with $ - {3 \over 2} $ $<$\nobreak\ $\delta$\nobreak\ $<$ - $ {1 \over 2} $,
hence in $ H_{3,-1} $ (not in $ H_{3,0} $), and analogously for the 
second time derivative with 3 replaced by 2. \par
\medskip
\noindent\underbar{Lemma 2.} \underbar{Hypothesis.} We give on
M$\times$I, I $\equiv$ $\lbrack$0,T$\rbrack$: \par
\smallskip
\noindent a) $\beta$ $\in$ $ \doublelow{ \cap \cr 1\leq k\leq 3 \cr} $C$
^{k-3}(I,H_{k,-1} $),\nobreak\ \nobreak\  h $\in$ $ \doublelow{ \cap \cr
2\leq k\leq 3 \cr} $ $ C^{k-3}(I,H_{k,1}) $ \par
\noindent b) N(n) $\geq$ N$ _0 $ $>$ 0 , N(n) $\in$ $  \doublelow{ \cap \cr
0\leq k\leq 2 \cr} C^{2-k}(I,H_{2+k,-1}) $ \par
\noindent c)  $ \bar  g(n) - e \in   \doublelow{ \cap \cr 1\leq k\leq 3 \cr}
C^{3-k}(I,H_{k,-1}), K(n) \in   \doublelow{ \cap \cr 0\leq k\leq 2 \cr}
C^{2-k}(I,H_{k,0} $), \par
\noindent with moreover $ \partial_ tK_n $$\in$ $ \doublelow{ \cap \cr 0\leq
k\leq 1 \cr}\allowbreak C^{1-k}(I,H_{k,1}) $ . \par
\smallskip
\noindent d) The Cauchy data (independent of n) such that \par
\smallskip
\noindent$ g_{ij}(n+1)(0,.) = \gamma_{ ij},\quad \gamma  - e \in  H_{3,-1},
\ \ \ K_{ij}(n+1)(0,.) = k_{ij} \in  H_{2,0} $, \par
\smallskip
\noindent and the metric $\gamma$ is uniformly equivalent to the given metric e.
\par
\noindent The data $ \hat  \partial_ 0K_{ij}(n+1)(0,.) = \dot  k_{ij} $,
determined by the equation $ R_{ij}(0,.) = 0 $, 
belong to $ H_{1,1} $. \par
\smallskip
\noindent\underbar{Conclusion.} The Cauchy problem for the system (1$ _n $),
(2$ _n $) has a unique 
solution such that \par
\smallskip
\centerline{$ \bar  g(n+1) - e \in   \doublelow{ \cap \cr 1\leq k\leq 3 \cr}
C^{3-k}(I,H_{k,-1}), K(n+1) \in   \doublelow{ \cap \cr 0\leq k\leq 2 \cr}
C^{2-k}(I,H_{k,0}) $ }
\smallskip
\centerline{$ \partial_ tK(n+1) \in  \doublelow{ \cap \cr 0\leq k\leq 1 \cr}
$C$ ^{1-k}(I,H_{k,1}). $}
\medskip
\noindent There exists an interval $ I_1= \lbrack 0,T_1\rbrack $, 0 $<$ T$ _1
$ $\leq$ T, and a number $\epsilon >$ 0 
depending only on the norms of the given quantities in their respective 
spaces such that $ \bar  g $(n+1) is uniformly equivalent to $\gamma$ on
M$\times I_1 $. \par
\smallskip
\noindent\underbar{Proof.} The first weighted energy estimate needed to prove
the lemma is
obtained for $ DK $(n+1) (denoted $ K $ to simplify the writing) and
$\partial_t K $  by 
multiplying its wave equation by $\sigma^2\nabla_ 0K^{ij} $, after
checking that under the 
hypothesis made on the n$ ^{th} $ iterates we have $\sigma P_{ij}(n) $
$\in$ C$ ^0(I,L^2) $. The 
estimates for $\partial_t DK $ and D$ ^2K $ are obtained after derivation
of the equation in 
the space directions. We obtain their boundedness in $ H_{0,2} $ because 
$ \sigma^ 2DP_{ij} \in  C^0(I,L^2).  $The estimate for $ \partial^ 2_{tt}K $
is obtained only in H$ _{0,1} $ because 
only $\sigma \partial_t P_{ij} $ $\in$ $ C^0(I,L^2) $, due to the term in
D$ ^2\partial^ 2_{tt}N  $. \par
\noindent The estimate for K can be obtained by integration in t from the
estimate 
of $\partial_t K$:  it implies $ K - k \in  C^0(I,H_{0,1} $, hence $ K $
$\in$ C$ ^0(I,H_{0,0} $) if it is 
so of k. \par
\noindent The property $ \partial_ t\bar  g(n+1) $ $\in$ $ C^0( $I,H$ _{2,0})
$, $\partial$ $ ^2_{tt}\bar  g(n+1) \in  C^0(I,H_{1,1}) $, $\partial$ $
^3_{ttt}\allowbreak\bar  g(n+1) \in \ \   $ 
C$ ^0(I,H_{0,1}) $ results from the equation (1$ _n) $. By integration with
respect to 
t we obtain $ \bar  g (n+1) - \gamma$ $\in$ C$ ^0(H_{2,0})$,  hence $ \bar 
g(n+1) $ $-$ e $\in$ $ C^0(I,H_{2,-1}) $ by the 
hypothesis on $\gamma$. To prove that it belongs  to C$ ^0(I,H_{3,-1} $) one
must use the 
third order equation for $ \bar  g(n+1)  $ deduced from (1$ _n), (2_n) $. \par
\smallskip
\noindent The convergence of the iterates in a small enough time interval is 
proved analogously as in the case of a compact M and gives the following 
theorem. \par
\smallskip
\noindent{\bf Theorem.} The system (1), (2), (3) with Cauchy data $\gamma$ and
k on the manifold 
(M,e), euclidean at infinity with given $\beta$ and h on M$\times
\lbrack$0,T$\rbrack$ satisfying the 
hypothesis spelled out in lemmas 1 and 2, has one and only one solution
($ \bar  g $,K,N) 
on M$\times$J, J $\equiv$ $\lbrack$0,$\ell \rbrack$ a sufficiently small
subinterval of I,  if the Cauchy 
data are such that $\gamma$ - e $\in$ H$ _{3,-1} $ and is uniformly equivalent
to e, k $\in$ H$ _{2,0}
$, $ \dot  k $ $\in$ H$ _{1,1} $. The solution belongs to the functional spaces
indicated for the 
iterates in the lemmas and is such that N $>$ 0 and $ \bar  g $ is uniformly 
equivalent to $\gamma$. \par
\smallskip
\noindent\underbar{Remark.} The estimates show in fact that $ \bar  g_t -
\gamma $ is such that for each t we 
have \par
\smallskip
\centerline{$ \bar  g_t - \gamma  \in  H_{3,0}, $}
\smallskip
\noindent i.e., it has a stronger asymptotic fall off than $\gamma$ - e 
or $ \bar g_t - e  $; this 
property is related to the A.D.M.$^{3}$ theorem of mass conservation.
\bigskip
\noindent SOLUTION OF THE FULL EINSTEIN EQUATIONS. \par
\smallskip
\noindent We suppose that we have solved the reduced equations, that is \par
\smallskip
\centerline{$ \Omega_{ ij} + N \bar  \nabla_ i\partial_ j(H - h) = \Theta_{
ij} $}
\noindent and \par
\smallskip
\centerline{$ R_{00} - N \partial_ 0(H - h) = \rho_{ 00} $}
\medskip
\noindent We will show that the solution obtained satisfies the original
Einstein 
equations if the source $\rho$ is such that the associated stress energy
tensor 
satisfies the conservation laws
\smallskip
$$ \nabla_ \alpha T^{\alpha \beta}  = 0  $$
where (8$\pi$G = c =1)\par
\smallskip
\centerline{$ T_{\alpha \beta}  \equiv  \rho_{ \alpha \beta}  - {1 \over 2}
g_{\alpha \beta} \rho $}
\smallskip
\noindent and the initial data $\gamma$ and k satisfy the constraints: 
\par
\centerline{$ S^0_0 = T^ 0_0,\ \ \ S^0_i \equiv  T^ 0_i,\ \ \ {\rm for}
\ \ \ \    t = 0 $}
\smallskip
\noindent with moreover \par
\smallskip
\centerline{tr\thinspace k $\equiv$ H$\mid$ $ _{t=0} $ = h$\mid$ $ _{t=0} $}
\smallskip
\noindent while the initial values of $\hat \partial _0K $ are chosen so that
\par
\smallskip
\centerline{$ R_{ij}\mid_{ t=0} = \rho_{ ij}\mid_{ t=0} $,}
\smallskip
\noindent i.e., the Einstein equations are satisfied initially. \par
\smallskip
\noindent The proof follows with $ A_{\alpha \beta}  \equiv  R_{\alpha
\beta}  - \rho_{ \alpha \beta} $, H $-$ h $\equiv$ X. We will use the 
Bianchi identities $\nabla_\alpha S^{\alpha \beta}  \equiv  0$  where \par
\smallskip
\centerline{$ S_{\alpha \beta}  = T_{\alpha \beta} ,\ \ \ \ \ S_{\alpha
\beta}  \equiv  R_{\alpha \beta}  - {1 \over 2} g_{\alpha \beta} R $}
\medskip
\noindent For any covariant symmetric space time 2-tensor A we find by using
the 
values of the $\omega$'s  that we have the identities \par
\smallskip
\centerline{$ \nabla_ 0A_{ij} \equiv  \hat  \partial_ 0A_{ij} + N
K^h_{(i}A_{j)h} $ $ - N^{-1}\partial_{ (i}N A_{j)0} $}
$$\nabla_iA_{j0} \equiv \bar \nabla _iA_{j0} +N^{-1}K_{ij}A_{00} - N^{-1}
A_{0j}\partial_iN + NA_{hj}K^h_i$$\par
$$\nabla_i\partial_jX \equiv \bar \nabla_i\partial_jX 
+ N^{-1}K_{ij}\partial_0X$$\par
\smallskip
\noindent A straightforward computation using these identities shows 
that the reduced equations can also be 
written: \par
\smallskip
$$ \nabla_ 0A_{ij} - \nabla_{ (i}A_{j)0} +  
N \nabla_ i\partial_ jX + K_{ij}\partial_0X = 0  $$
\smallskip
\noindent and \par
\centerline{$ A_{00} - N \partial_ 0X = 0 $}
\smallskip
\noindent The Bianchi identities together with the conservation laws give the 
conservation equations: \par
\smallskip
\centerline{$ \nabla_ \alpha \Sigma^{ \alpha \beta}  = 0,\ \ \ \ \ {\rm with}
\ \ \ \ \Sigma^{ \alpha \beta}  \equiv  S^{\alpha \beta}  - T^{\alpha \beta}
$}
\medskip
\noindent In our frame we have the identities \par
\smallskip
\centerline{$ \Sigma^{ 00} \equiv  {1 \over 2} (A^{00} + N^{-2} g^{ij}A_{ij})
$}
\smallskip
\centerline{$ \Sigma^{ ij} \equiv  A^{ij} +  {1 \over 2} g^{ij} (N^2 A^{00} -
g^{hk}A_{hk}) $}
\smallskip
\noindent The conservation equation with $\beta$ = 0 implies therefore \par
\smallskip
\centerline{$ {1 \over 2} (\nabla_ 0A^{00} + N^{-2}g^{ij}\nabla_ 0A_{ij}) +
\nabla_ iA^{i0} = 0 $.}
\smallskip
\noindent Hence, when the reduced equations are satisfied (recall that
$\nabla_0 N = 0) $, \par
\smallskip
\centerline{$ -  N^{-2} \nabla_ 0\partial_ 0X + g^{ij}\nabla_
i\partial_ jX + N^{-1}H\partial_0 X = 0. $}
\smallskip
\noindent This wave equation for X implies $ X $ = 0 on M$\times$I if $ X $
and $ \partial_ 0X $ vanish 
initially. We have then also $A_{00}$ = 0\par
\noindent We can now proceed in a manner analogous to $ ^6 $. We 
differentiate the conservation equation with $\beta$ = j and use the Ricci 
identity to obtain \par
\smallskip
\centerline{$ \nabla_ 0\nabla_ 0A^{0j} + \nabla_ i\nabla_ 0A^{ji} + R^{\ \
j}_{0i\ \lambda} A^{\lambda i} - R_{0\lambda} A^{j\lambda} 
-{1\over 2} \nabla^ j\nabla_ 0A^{i}_{\ i} +
{1\over 2} N^2 \nabla^ j\nabla_ 0A^{00}$= 0.}
\smallskip
\noindent Using the reduced equations now with $A_{00} = X = 0$, we find\par
\smallskip
\centerline{($ \nabla_ 0\nabla_ 0 - N^2\nabla_ i\nabla^ i)A^{j0} - 
N^2(\nabla_ i\nabla^ jA^{i0}- \nabla^ j \nabla_iA^{i0}) +
{1\over 2} N^2 \nabla^ j\nabla_ 0A^{00} \sim  0 $}
\smallskip
\noindent where $\sim$ 0 means modulo linear terms in $ A^{i\lambda} $. Using
again the Ricci identity 
and the conservation law with $\beta =0$,  we find \par
\smallskip
\centerline{$-\emptysq$ $ A^{j0} $ $\equiv$ ($ N^{-2}\nabla_ 0\nabla_ 0 -
\nabla_ i\nabla^ i)A^{j0} \sim  0 $}
\smallskip
\noindent where  $\sim$ 0 denotes terms linear in $ A^{i\lambda} $ and
its derivatives of order $\leq$ 1.  
We deduce then from the first reduced equation: \par
\smallskip
\centerline{$ \emptysq \nabla_ 0A_{ij} \sim  0 $}
\smallskip
\noindent where $\sim$ 0 is modulo linear terms in $ A^{i\lambda} $ and their
derivatives of order 
$\leq$ 2. We have obtained for $ A_{i\lambda} $ 
a strictly hyperbolic system which is linear and homogeneous. The 
uniqueness theorem of Leray$^8$ for solutions of such systems shows that we 
have $A_{i \lambda}$ = 0  on M$\times$I if the corresponding initial data
are zero.\par 
\noindent The set of initial data in our problem  
 are the values on M$ _0 $ of X, $ \partial_ 0X, $ $ A_{i0}  $, $
\partial_ 0A_{i0} $ , $ A_{ij}, \partial_ 0A_{ij} , \partial^ 2_{00}A_{ij} $. 
These values can be proved to be all zero for a solution of the reduced 
equations if $ (H-h)\mid_ t = 0 $ and the Einstein equations are satisfied on
M$ _0  $. Our proof is complete.
\bigskip
\centerline{II. HARMONIC TIME, ALGEBRAIC GAUGE}
\smallskip
\noindent HYPERBOLIC SYSTEM FOR $ \bar  g $, K, N. \par
\smallskip
\noindent We shall eliminate at the same time the third derivatives of N and
the 
second derivatives of H in the second order equation for $ K $ by a 
particular choice of N which we now explain (cf. this elimination in the 
case of zero shift in Choquet Bruhat and Ruggeri$^6$). We compute \par
\smallskip
\centerline{$ \hat  \partial_ 0\bar  \nabla_ j\partial_ iN \equiv  (
{\partial \over \partial t} - {\cal L}_\beta )\bar  \nabla_ j\partial_ iN $}
\smallskip
\noindent We find at once \par
\smallskip
\centerline{$ {\partial \over \partial t} \bar  \nabla_ j\partial_ iN \equiv 
\bar  \nabla_ j\partial_ i {\partial N \over \partial t} - {1 \over 2}
g^{kl}(\bar  \nabla_{ (i} {\partial \over \partial t} g_{j)l} - \bar  \nabla_
l {\partial \over \partial t} g_{ij})\allowbreak \partial_ kN $}
\smallskip
\noindent and we find an anlogous formula when the operator
$\partial$/$\partial$t is replaced by ${\cal L}_\beta $ by 
using the property \par
\smallskip
\centerline{$ \partial_ i{\cal L}_\beta N \equiv  {\cal L}_\beta \partial_ iN
\equiv  0 $}
\smallskip
\noindent which leads to \par
\smallskip
\centerline{$ {\cal L}_\beta\bar  \nabla_ j\partial_ iN - \bar  \nabla_
j\partial_ i{\cal L}_\beta N \equiv  - \partial_ hN (\bar  \nabla_ j\bar 
\nabla_ i\beta^ h - \beta^ l\bar  R^{\ \ \ h}_{lji} $)}
\smallskip
\centerline{$  \equiv  -  {1 \over 2}  g^{kl}(\bar  \nabla_{ (i}{\cal
L}_\beta g_{j)l} - \bar  \nabla_ l{\cal L}_\beta g_{ij})\allowbreak \partial_
kN $}
\noindent$   $ \par
\noindent Finally \par
\smallskip
\centerline{$ \hat  \partial_ 0\bar  \nabla_ j\partial_ iN \equiv  \bar 
\nabla_ j\partial_ i\partial_ 0N \allowbreak  + {1 \over 2} (\bar  \nabla_{
(i}  K_{j)l} - \bar  \nabla_ l  K_{ij})\allowbreak \partial^ lN $.}
\smallskip
\noindent The third order terms in N and second order terms in H in the
second 
order equation for K can therefore be written under the following form \par
\smallskip
\centerline{$ C_{ij} \equiv $ $ - $N$ ^{-1} \bar  \nabla_ j\partial_
i(\partial_ 0N + N^2H) $}
\smallskip
\noindent We satisfy the condition $ C_{ij} = 0  $ by imposing on N the 
differential equation \par
\smallskip
\noindent (N$^\prime )\ \ \ \ \ \ \ \ \ \ \ \ \ \ \ \ \ \ \ \ \ \ \ \
\partial_ 0N + N^2H = 0 $ \par
\smallskip
\noindent The second order equation for K reads then as a wave type non linear
system: \par
\medskip
\noindent (K$^\prime$)\nobreak\ \nobreak\ \nobreak\ \nobreak\ \nobreak\ 
\nobreak\
\nobreak\ \nobreak\ \nobreak\ \nobreak\ \nobreak\ \nobreak\ \nobreak\
\nobreak\ \nobreak\ \nobreak\ \nobreak\ \nobreak\ \nobreak\ \nobreak\
\nobreak\ N$  \hat  {\emptysq} K_{ij} = N Q_{ij} + \Theta_{ ij} $ \par
\smallskip
\noindent where we now set \par
\smallskip
\centerline{$ \hat  {\emptysq} K_{ij} \equiv  - N^{-2}\hat  \partial_ 0\hat 
\partial_ 0K_{ij} + \bar  \nabla^ h\bar  \nabla_ hK_{ij} $}
\smallskip
\centerline{$N  \eqalignc{ Q_{ij}\equiv  - K_{ij}\partial_ 0H  +
2g^{hm}K_{m(i} \hat  \partial_ 0K^{ }_{j)h} + 4Ng
^{hl}g^{mk}K_{lk}K_{im}K_{jh} \cr} $}
\smallskip
\centerline{+ $(2 \bar  \nabla_{ (i}K_{j)l})\allowbreak \partial^ lN  - 2H 
N^{-1}\partial_ iN \partial_ jN - 2\partial_{ (i}N \partial_{ j)}H $}
\smallskip
\centerline{$- 3 \partial_ hN \bar  \nabla^ hK_{ij} - K_{ij}\bar  \nabla^
h\bar  \nabla_ hN  - N^{-1}K_{ij}\partial^ hN \partial_ hN $\nobreak\ + $
N^{-1}K_{h(i}\partial_{ j)}N \partial^ hN  $}
\smallskip
\centerline{+ ($ \bar \nabla _{ (i}\partial^ hN) K_{j)h} $ +
$2N \bar R^h_{\ ijm}K^m_h + N \bar R_{m(i}K^m_{j)} -  2H \bar \nabla 
_ j\partial_ iN $}
\smallskip
\centerline{$ \Theta  _{ij} \equiv   \hat  \partial_ 0\rho_{ ij}  - 
\bar \nabla _{ (i} \rho^{\ 0}_{j)}   $}
\medskip
\noindent The equation (N$'$) expresses that the time coordinate is harmonic,
namely: \par
\smallskip
\centerline{$ \nabla^ \alpha \partial_ \alpha x^0 \equiv  N^2\omega^ 0_{00} -
g^{ij}\omega^ 0_{ij} = 0. $}
\smallskip
\noindent Using the expression for $H$ we see that this equation reads \par
\smallskip
\centerline{$ \hat  \partial_ 0log\lbrace N/( {\rm det}  \bar  g)^{1/2}
$$\rbrace$ = 0}
\smallskip
\noindent We find, generalizing the result obtained by Choquet-Bruhat and 
Ruggeri$^6$ in the 
case of zero shift, that the general solution of this equation is \par
\smallskip
\centerline{$ N = \alpha^{-1}  {\rm (det}\bar  g^{1/2}) $}
\smallskip
\noindent where $\alpha$ is an arbitrary tensor density such that \par
\smallskip
\centerline{$ \hat  \partial_ 0\alpha  = 0. $}
\smallskip
\noindent This equation is a linear first order partial differential equation
for $\alpha$ 
on space time, depending only on $\beta$. A possible choice if $\beta$ does
not depend 
on t is to take $\alpha$ independent of t and such that $ {\cal
L}\allowbreak_ \beta \alpha  = 0. $ \par
\noindent The above choice of N is called algebraic gauge. \par
\smallskip
\noindent If we replace N by the value obtained above  in the second order
equation 
for $ K $, we obtain a quasi diagonal system with principal part the wave 
operator, with terms depending on $ \bar  g $ and its derivatives of order
$\leq$ 2. This 
system reduces to a third order hyperbolic system when we replacing $ K $ by 
$-(2N)^{-1}\hat  \partial_ 0\bar  g $. The local existence theorem of Leray$^8$
for the solution of 
hyperbolic systems gives immediately the local in time existence of 
solutions of this reduced system, in local in space Sobolev spaces. \par
\noindent It can be proved that a solution of the reduced system on M$\times$I
is a 
solution, in algebraic gauge, of the full Einstein equations if the initial 
data satisfy the Einstein equations on M$ _0 $ (cf. the proof in the case of 
zero shift in$ ^6 $). \par
\medskip
\smallskip
\noindent FIRST ORDER SYSTEM (vacuum). \par
\smallskip
\noindent The preceding results can be extended without major change to
dimensions 
greater than 4. We will show that in dimension 4 a solution of the vacuum 
Einstein equations, together with the harmonic time gauge condition, 
satisfies a first order symmetric system, hyperbolic if $ \bar  g $ is
properly 
riemannian and N$^2 >$ 0. Such a system could be useful to establish a
priori 
estimates relevant to global problems. It is of great importance for numerical 
computations, because symmetric hyperbolic systems occur in many areas of 
mathematical physics, in particular in fluid dynamics, and codes have been 
developed to study such systems. \par
\smallskip
\noindent We have obtained for the unknowns $ \bar  g $, K, N the equations
\smallskip
$$ \hat  \partial_ 0g_{ij} = - 2NK_{ij} \leqno (1) $$
$$ \partial_ 0N = - N^2H \leqno (2) $$
$$  \hat  {\emptysq} \ K_{ij} \equiv  Q_{ij} \leqno (3) $$
\par
\noindent To obtain a first order system we take as additional unknowns: \par
\smallskip
$ \hat  \partial_ 0K_{ij} = M_{0ij} $, $ \bar  \nabla_ hK_{ij} = M_{hij} $, $
\partial_ iN = $ $ N_i $, $ \hat  \partial_ 0\partial_ iN = N_{0i} $, $ \bar 
\nabla_ j\partial_ iN = N_{ji}. $ \par
\smallskip
\noindent We take as equation (3$'$) \par
$$\hat  \partial_ 0K_{ij} = M_{0ij} \leqno (3') $$ \par
\noindent The equation (3) gives
$$ \hat  \partial_ 0M_{0ij} - N^2\bar  \nabla^ hM_{hij}= - N^2 Q_{ij} \leqno
(4) $$
\par
\noindent In three space dimensions the riemann tensor is a linear function of
the Ricci tensor:
\smallskip
$$ \bar  R_{lijm} \equiv  g_{lj}\bar  R_{im} + g_{im}\bar  R_{jl} -
g_{ij}\bar  R_{lm} - g_{lm}\bar  R_{ij}- {1 \over 2} (g_{lj}g_{im}-
g_{ij}g_{lm})\bar  R $$
\par
\noindent Using the equation $ R_{ij} = 0$  we have \par
\smallskip
\centerline{$ \bar  R_{ij} = N^{-1}M_{0ij} - HK_{ij} + 2K_{ih}K^h_j +
N^{-1}N_{ji} $}
\smallskip
\noindent We then can write $N^2 Q_{ij} $ as a polynomial in the unknowns, N$
^{-1} $  and $ g^{ij} $. \par
\smallskip
\noindent We prove the following lemma for instance by using the commutativity
$ \hat  \partial_ 0\partial_ i = \partial_ i\hat  \partial_ 0 $ on components
of tensors and the value of $ \hat  \partial_ 0\bar\Gamma^ h_{ij} $ 
in terms of K:
\par
\smallskip
\noindent\underbar{Lemma.} For an arbitrary covariant vector u$ _i $ we have
\par
\smallskip
\centerline{$ \hat  \partial_ 0\bar  \nabla_ hu_i = \bar  \nabla_ h\hat 
\partial_ 0u_i + u_l\lbrace\bar  \nabla_ h(NK^l_i) + \bar  \nabla_ i(NK^l_h) -
\bar  \nabla^ l(NK_{ih})\rbrace $}
\smallskip
\noindent and an analogous formula for tensors with additional terms for each
index. \par
\smallskip
\noindent Using this lemma we see that $ M_{0ij} $ and $ M_{hij} $ must satisfy
the equations
\smallskip
$$ \hat  \partial_ 0M_{hij} - \bar  \nabla_ hM_{0ij} = N K_{l(j}(M^{\ \
l}_{hi)} + M^{\ \ l}_{i)h} - M^{l\ \ }_{\ i)h})  
+  K_{l(j} (K^l_{i)}N_h + N_{i)}K^l_h - K_{i)h}N^l). \leqno (5)$$ \par
\smallskip
\noindent On the other hand by definition
\smallskip
$$ \hat  \partial_ 0N_i = N_{0i}  \leqno (6) $$
\par
\noindent while $ N_{hi} $ and N$ _{0i} $ must satisfy
\smallskip
$$ \hat  \partial_ 0N_{hi}- \bar  \nabla_ hN_{0i} = NN_l(M^{\ \ l}_{hi} + M^{\
\ l}_{ih} - M^l_{\ ih})  
+  N_l (K^l_iN_h + K^l_hN_i - N^lK_{ih}).  \leqno (7)$$ \par
\smallskip
\noindent Now we deduce the value of $ \hat  \partial_ 0\hat  \partial_ 0 N_i
\equiv  \hat  \partial_ 0N_{0i} $ from (2) and from the Einstein 
equation \par
\smallskip
\centerline{$ R^0_0 \equiv  -\lbrace N^{-1}\bar  \nabla^ h\bar  \nabla_ hN -
K_{ij}K^{ij} + N^{-1}\partial_ 0H\rbrace  = 0 $}
\smallskip
\noindent Indeed these two equations imply that N satisfies the following 
inhomogeneous wave equation \par
\smallskip
\centerline{$ \partial_ 0\partial_ 0N - \allowbreak N^2\bar  \nabla  ^h\bar 
\nabla_ hN = - N^3K_{ij}K^{ij} + 2N^3H^2 $}
\smallskip
\noindent Hence by differentiation, use of the Ricci formula and the 
definitions of
$ N_{hi} $ and 
N$ _{0i} $, we find \par
\medskip
\noindent (8)\nobreak\  $$ \hat  \partial_ 0N_{0i}- \allowbreak N^2\bar 
\nabla^ hN_{hi}= - \bar  R^h_iN_h N^2 + 2NN_iN^h_{\ h} - 2N^3K_{hl}M^{\ hl}_i
$$
$$- 3N^2N_iK_{hl}K^{hl}+ 4N^3HM^{\ h}_{i\ h} +6 N^2 H^2 N_i. $$ 
\medskip
\noindent We use again the equation $ R_{ij}= 0 $ to replace $ \bar R_{ij} $
by its value in terms of 
the unknowns. \par
\smallskip
\noindent We have obtained a first order system (equations 1, 2, 3$'$, 4, 
5, 6, 7, 8) 
in all the unknowns. The right hand sides are polynomial in the unknowns,  
$ g^{ij} $ and N$ ^{-1} $. They do not depend on their derivatives. The left
hand sides 
are linear operators on the all the unknowns. Their coefficients depend on 
these unknowns, and not on their derivatives except for the derivatives of 
$ \bar  g. $ In order to obtain a covariant quasilinear first order system for
all 
the unknowns we can, for instance, introduce on M an a priori given metric e which may depend 
on t but is such that \par
\smallskip
\centerline{$ \hat  \partial_ 0e_{ij} = 0. $}
\smallskip
\noindent We denote by D the covariant derivative in the metric e. We
introduce the 
additional unknown $ G_{hij} \equiv  D_hg_{ij} $ and deduce from (1) the
equation:
\smallskip
$$ \hat  \partial_ 0G_{hij} = -2\lbrace N_hK_{ij} + N(M_{hij} +
S^m_{h(i}K_{j)m})\rbrace \leqno (9) $$
\par
\noindent where the tensor S, the difference of the connections of 
$ \bar  g $ and e, is given by \par
\smallskip
\centerline{$ S^m_{ij} =  {1 \over 2} g^{mh}(G_{(ij)h} - $ $ G_{hij} $)}
\smallskip
\noindent We have now obtained a quasi linear first order system for all the
unknowns. 
Its characteristic matrix, obtained by replacing in the principal matrix 
the operator $\partial$ by a covariant vector $\xi$, is constituted of blocks
around the 
diagonal, some reduced to one element $\xi _0 $, and some 4$\times$4 matrices
with 
determinant $\xi ^2_0 - N^2\xi^ i\xi_ i $. The characteristics are the
light cone and the 
time axis. On the other hand the system can be symmetrized by multiplication 
with a matrix constituted of blocks around the diagonal equal to one 
element, 1, or the matrix ($ g^{ij} $). In other words the system is a first
order 
\underbar{symmetrizable hyperbolic system,} with domain of dependence
determined by 
the \underbar{light cone}. Known local existence theorems apply to such a
system. \par
\smallskip
\noindent\underbar{Remark.} It is possible to obtain a system whose right hand
sides are 
polynomial in the unknowns and $ g^{ij} $ by making a slightly different choice
of 
these unknowns, in particular by introducing $ a_i \equiv  N^{-1}\partial_ iN $
(cf.$ ^{13,14} $). \par
\smallskip
\noindent J.W.Y. acknowledges support from the National Science Foundation
of the USA, grants PHY-9413207 and PHY 93-18152/ASC 93-18152 (ARPA
supplemented).\par
\smallskip
\centerline{{\bf Bibliography}}
\smallskip
\noindent 1. A. Lichnerowicz, ``Probl\`emes Globaux en Mecanique Relativiste",
Hermann 1939. \par
\noindent 2. Y. (Foures)-Bruhat, J. Rat. Mechanics and Anal. {\bf 5} p.951-966,
1956. \par
\noindent 3. A. Arnowitt, S. Deser and C. Misner in ``Gravitation: An 
Introduction to Current Research", L. Witten ed. Wiley 1962. \par
\noindent 4. A. Fisher and J. Marsden, J. Math. Phys. {\bf 13} p.546-568, 1972.
\par
\noindent 5. Y. Choquet-Bruhat and J. W. York, ``The Cauchy Problem" in General
Relativity and Gravitation, A. Held ed. Plenum 1980. \par
\noindent 6. Y. Choquet-Bruhat and T. Ruggeri, Comm. Math. Phys. {\bf
89}, p.269-275, 1983. \par
\noindent 7. D. Christodoulou and S. Klainerman, ``The Global Nonlinear 
Stability of the Minkowski Space " Princeton 1992. \par
\noindent 8. J. Leray, ``Hyperbolic Differential Equations" 
I. A. S. Princeton 1952. \par
\noindent 9. J. W. York, ``Kinematics and Dynamics of General Relativity" in
Sources of Gravitational Radiation, L. Smarr ed., Cambridge 1979. \par
\noindent 10. T. Hughes, T. Kato and J. Marsden, Arch. Rat. Mech. Anal. 
{\bf 63} p.273-274, 1976.\par
\noindent 11. A. Majda, ``Compressible Fluid Flow and Systems of Conservation 
Laws in Several Space Variables", Springer 1984.  \par
\noindent 12. Y. Choquet-Bruhat and D. Christodoulou, Acta Mathematica {\bf 146}
 p.129-150, 1981. \par
\noindent 13. Y. Choquet-Bruhat and J. W. York, C.R. Acad. Sci. Paris {\bf 321
}, p. 1089-1095, 1995. \par
\noindent 14. A. Abrahams, A. Anderson, Y. Choquet-Bruhat and J. W. York, 
Phys. Rev. Letters {\bf 75}, p.3377-3381, 1995. \par
\bigskip
Y.C.B. Gravitation et Cosmologie Relativiste, Un. Paris VI,t. 22-12,
75252\par 
J.W.Y. Department of Physics and Astronomy, UNC, Chapel Hill, NC 27599-3255
\par
\bye